\documentclass{article}
\usepackage{nips14submit_e,times}
\usepackage{hyperref}
\usepackage{url}

\usepackage[labelfont=bf,labelsep=period]{caption}
\usepackage{amssymb}
\usepackage{amsmath} 
\usepackage{graphicx}

\usepackage[numbers,square,sort,compress,comma]{natbib}

\title{On the role of time in perceptual decision making}

\author{M\'at\'e Lengyel\\
Department of Engineering\\
University of Cambridge\\
Cambridge, CB2 1PZ, UK\\
\texttt{m.lengyel@eng.cam.ac.uk} \\
\And
\'Ad\'am Koblinger \\
Department of Cognitive Science \\
Central European University \\
Budapest, H-1023, Hungary \\
\texttt{kob314@gmail.com} \\
\And
Marjena Popovi\'c \\
Neuroscience Program\\
Brandeis University\\
Waltham, MA 02454, USA \\
\texttt{marjenap@brandeis.edu} \\
\And
J\'ozsef Fiser\\
Department of Cognitive Science \\
Central European University \\
Budapest, H-1023, Hungary \\
\texttt{fiserj@ceu.hu}
}

\newcommand{\esterr}{\varepsilon_\mathrm{x}}
\newcommand{\estunc}{\sigma^2_\mathrm{x}}

\nipsfinalcopy 

\begin{document}

\maketitle

\begin{abstract}
According to the dominant view, time in perceptual decision making is used for integrating new sensory evidence.  Based on a probabilistic framework, we investigated the alternative hypothesis that time is used for gradually refining an internal estimate of uncertainty, that is to obtain an increasingly accurate approximation of the posterior distribution through collecting samples from it.  In the context of a simple orientation estimation task, we analytically derived predictions of how humans should behave under the two hypotheses, and identified the \emph{across-trial correlation between error and subjective uncertainty} as a proper assay to distinguish between them.  Next, we developed a novel experimental paradigm that could be used to reliably measure these quantities, and tested the predictions derived from the two hypotheses.  We found that in our task, humans show clear evidence that they use time mostly for probabilistic sampling and not for evidence integration.  These results provide the first empirical support for iteratively improving probabilistic representations in perceptual decision making, and open the way to reinterpret the role of time in the cortical processing of complex sensory information. 
\end{abstract}

\section{Introduction}

Making decisions is one of the most fundamental aspects of cognition and accordingly, its behavioral, neural and computational bases have been investigated extensively \citep{Edwards1954, Smith2004, Ma2006}.  The prevailing framework of perceptual decision making maintains that time in decision making is used for collecting evidence about the stimulus for the decision \citep{Gold2007}.  This proposal is corroborated by the fact that with the progression of time, the error of and the uncertainty about the decision in behavioral studies of decision making steadily decrease until they reach an asymptote presumably set by the internal noise of the system \citep{Ratcliff2008}. These behavioral studies handle decision making typically as a deterministic process and investigate the simplest two-alternative-forced-choice version of the problem yielding insights that are difficult to generalize to multiple options and estimation tasks (implying a continuum of options) abundant in real life \citep{Churchland2012, Tsetsos2010}.  

In recent years, a prominent alternative framework of perception and cognition emerged based on the concept of probabilistic representations and computation \citep{Knill1996}.  According to this view, humans and animals code not only a single value of the sensory input, but multiple values with their subjective uncertainty about those values, and thus representation and computation in the cortex is based on probability distributions \citep{Pouget2003}.  This proposal was supported by a number of studies finding that human and animal behavior in different tasks could be best described as if the participants were aware and used probabilistic representations to achieve their goals (approximately) optimally  \citep{Ernst2002, Kording2004,Courville2006, Pouget2013}, and gained further support from recent neurophysiological findings \citep{berkes11, Yang2007}. However, the probabilistic framework aggravates the concern of generalization: the exact probabilistic inference necessary to reach a decision in a multiple-choice or estimation task is computationally clearly intractable, and the potential of appropriate approximations has not been explored to date.

We selected one particular approximation strategy -- Monte Carlo sampling -- for probabilistic computation \citep{Fiser2010}, and explored the consequences of this framework on the role of time in decision making. Approximating probabilistic representations and computations by sampling means that time is needed to collect samples sequentially from the posterior distribution even when no new evidence enters the system and thus the posterior itself remains static.  This provides a new hypothesis with a novel role of time in sensory systems that complements evidence integration:  time is used for iteratively refining an approximation of the true posterior.  Thus, the goal of the present paper is twofold.  First, to find evidence at the earliest possible stage of visual information processing for probabilistic computations with iterative approximation.  Second, to establish whether evidence integration or probabilistic sampling dominates decision making in naturalistic estimation tasks.  

Our paper presents the following three new contributions.  First, we established a mathematical model of decision making under evidence integration (EI) and probabilistic sampling (PS), and by analytical derivation, we identified the measure of the  \textit{trial-by-trial correlation between error and uncertainty} as an appropriate measure to distinguish whether participants are predominantly occupied with EI or PS (Section 2).  Second, we developed a novel behavioral test method, which can be used for measuring immediate error and subjective uncertainty simultaneously on a trial-by-trial basis within $\sim$700 msec after stimulus presentation (Section 3).  Finally, we measured the  correlation between error and subjective uncertainty in this modified orientation estimation task as a function of stimulus presentation time, and found that the pattern of this correlation confirms the predictions derived from the PS hypothesis (Section 4).  Thus, we conclude that probabilistic computations take place from the earliest levels of cortical visual information processing, and that in simple perceptual decision making tasks with static stimuli the majority of time is used for improving an internal representation of the obtained information and not for incorporating new external evidence. We discuss our results in the broader context of perceptual decision making in Section 5.

\section{Theoretical predictions for evidence integration and probabilistic sampling}

In the Supplementary Information (SI) we derive analytical formul\ae{} for how estimation error, uncertainty, and their correlation should vary over time\footnote{``Time'' here corresponds to ``presentation time'' in the experiments described in the following sections.} in a perceptual (eg.\ orientation) estimation task under two orthogonal scenarios:
\begin{description}
\item[Evidence integration (EI)] The posterior distribution (over possible orientations) is represented exactly at any moment, and time is used to collect more evidence about the external stimulus thus leading to a gradual refinement of the corresponding posterior, ``homing in'' on the true stimulus value (orientation). 
\item[Probabilistic sampling (PS)] The posterior itself does not change in time\footnote{This can be justified by evidence integration converging to a finite-width posterior due to fundamental limits on perception, with convergence being fast because the stimulus is static.}, but it is represented approximately by drawing successive samples from it as in Monte Carlo techniques \cite{mackay03}, and so time is used to collect more samples from this static posterior.
\end{description}

\begin{figure}
\centering \includegraphics[width=0.6\textwidth]{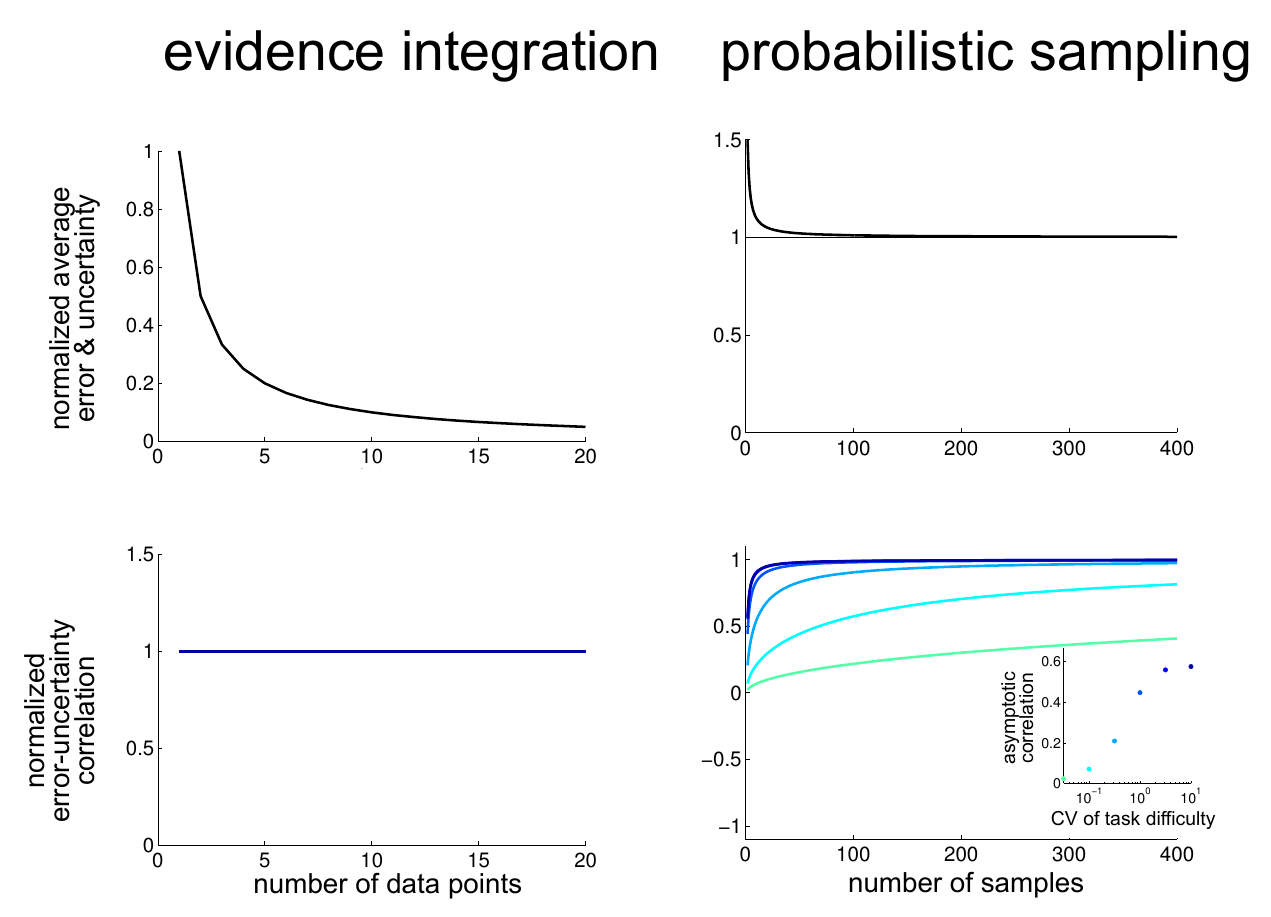}
\caption{\label{fig:analysis}{\bf Theoretical predictions.} Normalized average error and uncertainty (top) and the correlation between error and uncertainty (bottom) for evidence integration (EI, left) and probabilistic sampling (PS, right). Averages and correlations are computed across trials. Average error and uncertainty are normalised to their values obtained after the first data point for EI, and to the asymptotic value in the limit of infinite samples for PS. (An above-zero asymptote for EI can be obtained by introducing behavioural noise, see SI, or an asymptotically non-vanishing posterior variance.) Error-uncertainty correlation is normalised to its asymptotic value for both EI and PS, where the asymptotic value is obtained by using an exact representation of uncertainty, and is shown in the inset as a function of the coefficient of variation (CV) of the task difficulty (ie.\ the CV of the across-trial distribution of  posterior variance).}
\end{figure}

The assumptions underlying our derivations were the following:
\begin{itemize}
\item[--] The representation of uncertainty is self-consistent, in that uncertainty is predictive of the errors made in estimation (which we confirmed empirically, see below).
\item[--] Across trials, the posterior  changes its location and scale (mean and variance, which are both finite), but not its shape (higher moments). Importantly, while we show results here for a Gaussian posterior, the derivations themselves do not assume Gaussianity and work for any other shape with a known (and finite) kurtosis.
\item[--] Successive data points (for EI) or samples (for PS) are i.i.d. This means that the ``number of data points / samples'' referred to in the following should be understood as ``the effective number of independent data points /  samples'' which is proportional to the number of data points / samples and inversely proportional to some suitable measure of their statistical dependency (e.g.\ the total autocorrelation of samples when computing the Monte Carlo integral of a simple linear function under the posterior).
\item[--] For analytical tractability, there are two departures from the setting we tested experimentally: the variable over which the posterior needs to be represented is linear rather than circular, and estimation error is measured as squared rather than absolute error. We expect neither of these to impact our results qualitatively.
\end{itemize}
The main results of our derivations are the following (Fig.~\ref{fig:analysis}):
\begin{itemize}
\item[--] In both scenarios, the average error and uncertainty decrease roughly as 1 / time. This is interesting because a decrease in error with time has classically been taken as a hallmark of EI. We show that it can equivalently result from PS, without EI.
\item[--] The time course of the correlation between error and uncertainty distinguishes between EI and PS. In EI, it remains constant (or decreases if behavioural noise is considered), while in PS it may potentially transiently decrease, but eventually always increases even well after average error and uncertainty have already asymptoted.
\end{itemize}

\section{Measuring trial-by-trial error and uncertainty in an orientation estimation task}

\begin{figure}[t]
\includegraphics[width=0.6\textwidth]{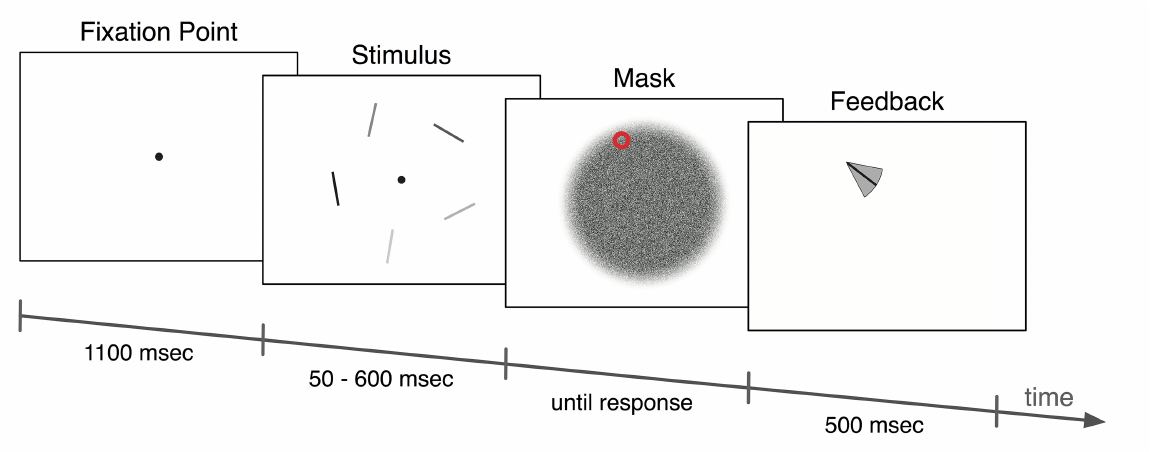}
\includegraphics[width=0.3\textwidth]{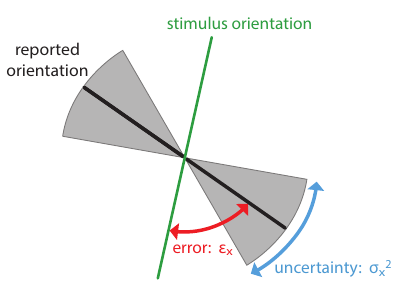}
\caption{\label{fig:expt}{\bf Experimental design.}
\textbf{A.}  In each trial, the subject made an orientation estimation judgement and provided information about the orientation and their subjective uncertainty by drawing a single stroke on a tablet.  See text for details. 
\textbf{B.}  Estimation error ($\esterr$) between the true and reported orientations, and level of uncertainty ($\estunc$) were the dependent variables in the experiment.} 
\end{figure}

In order to asses whether human decision making is dominated by EI or PS, we developed a new experimental paradigm (Figure~\ref{fig:expt}).  The basis of our paradigm was a standard orientation estimation task.  In each trial, subjects first saw a blank screen with a fixation dot for 1100 msec.  Next, a display appeared with a variable number (1--6, randomly chosen) of 1-degree-long line segments equidistant (with a randomly chosen rotation) around a circle (extending 7$^\circ$ of visual angle in diameter). Contrast levels were sampled randomly (without replacement within a display)  from \{10, 20, 30... 100\%\}, and orientations uniformly from 0$^\circ$-180$^\circ$. The display appeared for one of nine possible durations (``presentation time''): 50, 75, 100, 133, 167, 200, 300, 400, or 600 msec. After the display disappeared, a mask of random noise appeared with a small red circle identifying the position of one of the segments in the preceding display.  The subject's task was to report as quickly as they could their estimate of the orientation of the segment in the cued position simultaneously together with the subjective assessment of uncertainty in their estimate by drawing a line on a tablet with a stylus.  The orientation of the line indicated the estimated orientation of the line segment, while the length of the line corresponded to  subjective uncertainty.  (A longer line indicated less certainty to avoid the possible confound in measuring error-uncertainty correlations as shorter segments provide inherently less precision for orientation. ) After the subject responded, the mask and cue disappeared and a small segment appeared at the tested location with the orientation chosen by the subject and with a gray wedge around the line segment with a width (subtended angle) corresponding to the reported uncertainty. (The true orientation of the segment was not displayed.) This feedback display appeared for 500 msec, after which a new trial began. 

To enhance the quality of subjective uncertainty estimation, a scoring function was used to assess subjects' performance on each trial.   Subjects were instructed that their goal was to maximize their score which was calculated by combining the accuracy and certainty of their response.  As a scoring function, we used the log probability of the true stimulus orientation under a circular Gaussian (von Mises) distribution defined by the subject's response (segment orientation -- mean, wedge width -- concentration).  This scoring function can be maximised  if the subject's uncertainty report reflects their true subjective uncertainty which in turn is predictive of the errors they are making \cite{jaynes03}.   To prevent subjects from developing simple feedback-based strategies while keeping them alert, subjects received only grouped feedback after every 10 trials in the form of an average score.  Subjects completed 3-4 sessions of 900 trials across multiple days. To familiarize themselves with the procedure and to facilitate the precision of mapping from uncertainty to line length, prior to each test session subjects had a practice session with 50 trials during which they received feedback after every trial including the true orientation of the cued segment. Data from these trials was not included in the analyses.

\section{Results}

We collected data from a total of $N=5$ subjects, four of whom were naive while the last one was informed about the goal of the experiment.  We found no difference in performance between the naive and informed subjects confirming that the paradigm measured direct reactions of the subjects without much cognitive influence.

\subsection{Basic measures and controls}

\begin{figure}[t]
\includegraphics[width=\textwidth]{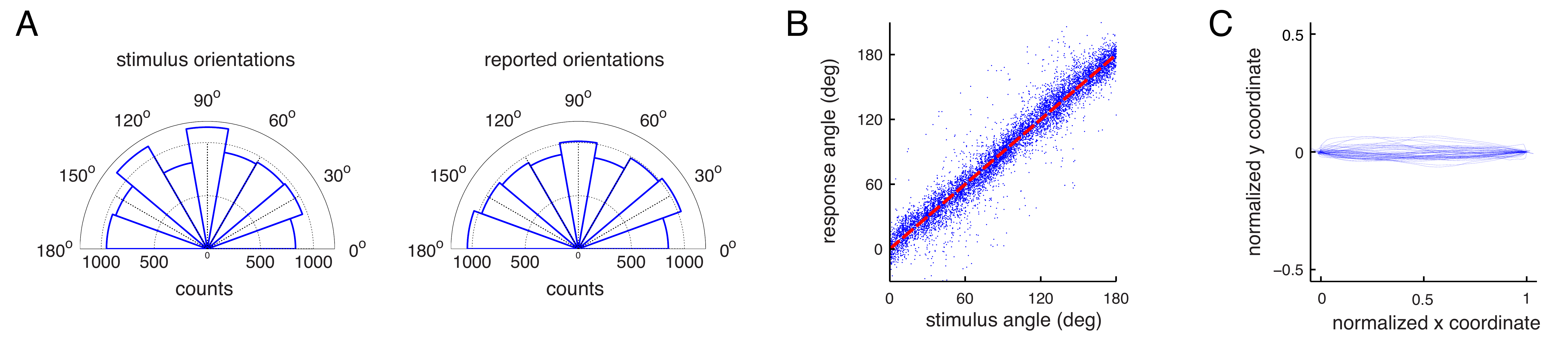}
\caption{\label{fig:control}{\bf Control measures.}
The experiment gives veridical trial-by-trial information about subjects' error and subjective uncertainty.
\textbf{A.} The distributions of the test line segments' true and reported orientation.
\textbf{B.} Trial-by trial correspondence between the line segments' true and reported orientation across all subjects and trials.
\textbf{C.} Trajectories of strokes for all subjects normalized (rotated and scaled) such that they go from (0,0) to (1,0).} 
\end{figure}

First, we checked whether in our paradigm we measured the relevant aspects of human performance. In order to measure the typical pattern of trial-by-trial error and uncertainty, the stimuli must cover the entire space of orientation, the subject's perception needs to follow the true stimuli, and response movements need to be ballistic.  Figure~\ref{fig:control} confirms that these requirements were fulfilled. The stimulus distribution was nearly uniform in the space of orientations, and the subjects' response was similar without any evidence for a bias in the cardinal directions (Fig 2A).  Subjects' judgement faithfully followed the true orientation of the target line segment (Pearson's r=0.97, p$<$10$^{-3}$) (Fig 2B), and their stroke was a straight line with average deviation from the straight line between the starting and endpoints below 3.3$\pm$0.2\% of the length of the stroke (Fig 2C).  In addition, we calculated the time profile of the strokes and found that subjects' mean duration of drawing was 450$\pm$110 msec (mean$\pm$s.e.) with standard deviation of  160$\pm$30 msec.  This suggests that subjects drew the line segments with a fast, single stroke without much fine-tuning, explicit cognitive deliberation, or modulation by different aspects of the task. 

\subsection{The representation of error and uncertainty}

\begin{figure}[t]
\includegraphics[width=\textwidth]{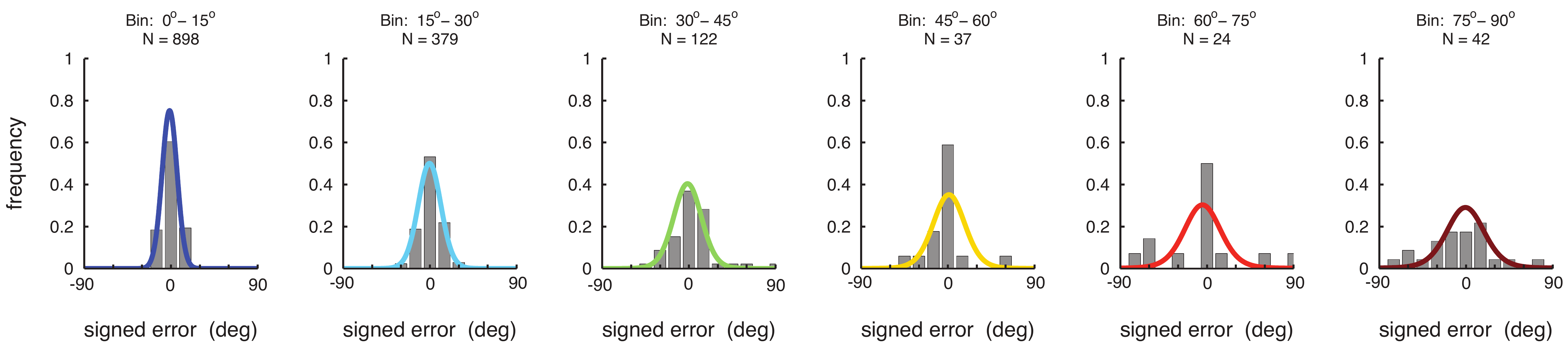}
\caption{\label{fig:erruncone}\textbf{Relation between error and subjective uncertainty in a single subject.}
Signed orientation estimation errors fitted with a circular Gaussian in trials binned according to the reported uncertainty (labels on top).} 
\end{figure}

Next, we tested whether subjects' uncertainty reports were predictive of their estimation errors.  Figure~\ref{fig:erruncone} shows a typical subject's result with trials binned by reported uncertainty and the resulting error histograms fitted with a circular Gaussian.  The error distribution remained centered at zero, but showed a clearly increasing spread as the subject's subjective report of uncertainty about the correctness of the trial increased.

Figure~\ref{fig:erruncall} shows the same fitted circular Gaussians to each subject together with the underlying scatter plots of (absolute) error vs.\ reported uncertainty.  Despite individual variations, each subject showed the same general relation of increasing uncertainty corresponding to steadily increasing error in their performance.  This suggests that subjects had a reliable representation of the quality of their perceptual information and faithfully reported this through their stroke. Thus, our experimental paradigm and response method successfully captured subjects' trial-by-trial error and uncertainty.

\subsection{The effect of task difficulty on error, uncertainty and their correlation}

We investigated how task difficulty affects subjects' error level, uncertainty and the correlation between the two.  As task difficulty increased either by increasing the number of line segments in the display or by decreasing the contrast of the target segment, both the error rate and uncertainty of the judgment increased significantly (Figure~\ref{fig:errunc_taskdiff}, 1st column, top two panels; the absolute value of Spearman's $\rho$ was between 0.60 and 0.95, with p$<$0.002 in all cases).  Expressed in terms of reaction times (RTs), we found the same trend: as RTs increased, presumably due to finding the trial more difficult either because of the nature of the sensory input (more segments, less contrast) or for some other reason (e.g.\ lapse of attention, or focusing on an irrelevant part of the display), both subjects' errors and level of uncertainty increased significantly (Figure~\ref{fig:errunc_taskdiff}, 1st column, bottom panel; Spearman's $\rho$$>$0.87, with p$<$0.001 in both cases).  Thus all three manipulations had a significant modulatory effect on both error and uncertainty of subjects' response.  The effect of contrast is particularly remarkable, as each line segment in a display had a different contrast level. Thus, the fact that the contrast level of the target line segment had an effect indicates that subjects possessed a multivariate representation of uncertainty, whereby they represented their uncertainty for multiple objects in the scene individually, rather than having just a single summary measure of uncertainty about the whole scene.

\begin{figure}[t]
\includegraphics[width=\textwidth]{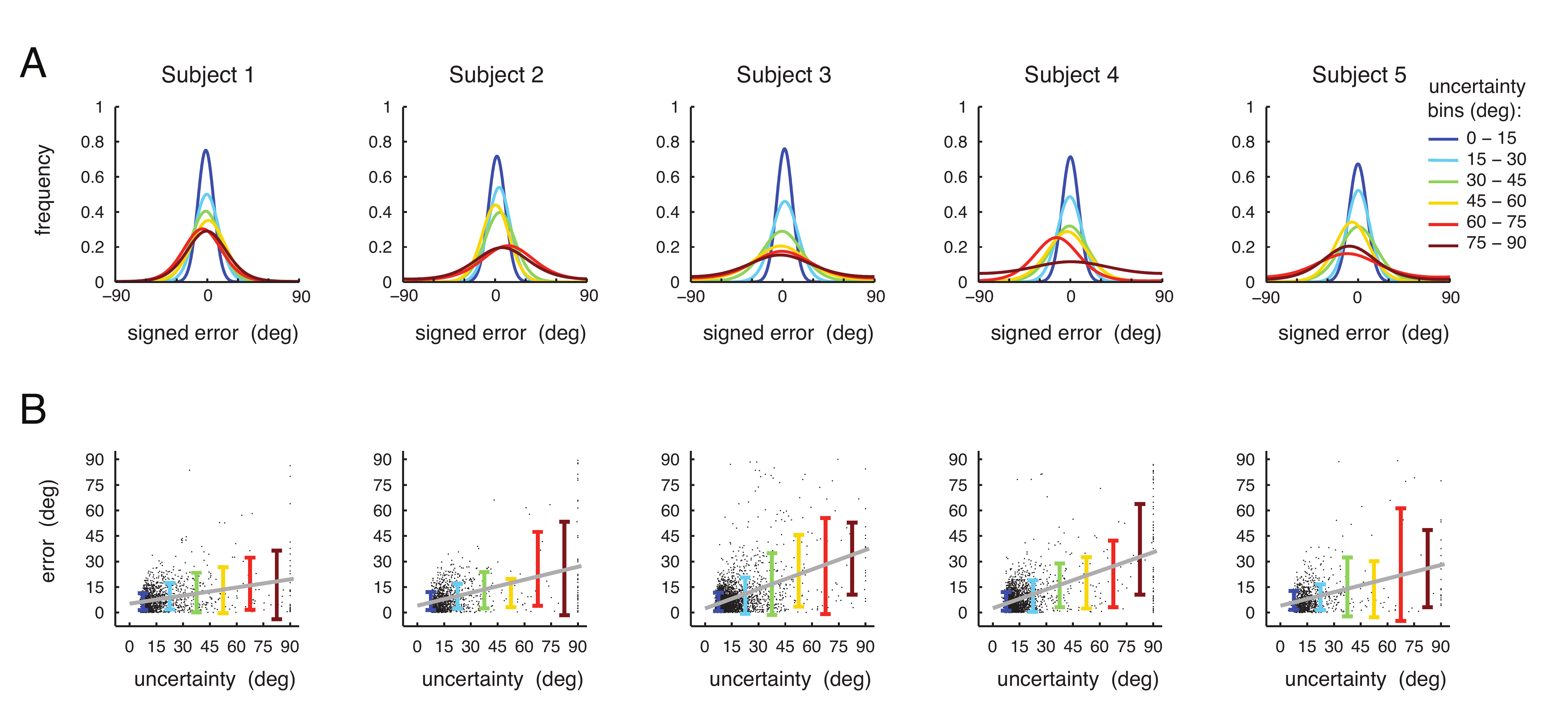}
\caption{\label{fig:erruncall}\textbf{Relation between error and subjective uncertainty across all subjects.}
Subjects' orientation estimation error changed according to their subjective uncertainty.
\textbf{A.}  Circular Gaussian fit of orientation estimation error histograms corresponding to different levels of reported uncertainty (as in Fig.~\ref{fig:erruncone}).
\textbf{B.}  Scatter plots of (absolute) error vs.\ uncertainty. Error bars show the mean$\pm$standard deviation for the data points in each uncertainty bin, gray lines represent linear regression fitted to the scatterplot. Note that it is the mean of absolute error plotted in panel \textbf{B} that is related to the standard deviation of signed error in panel \textbf{A}.} 
\end{figure}

\begin{figure}[t]
\includegraphics[width=\textwidth]{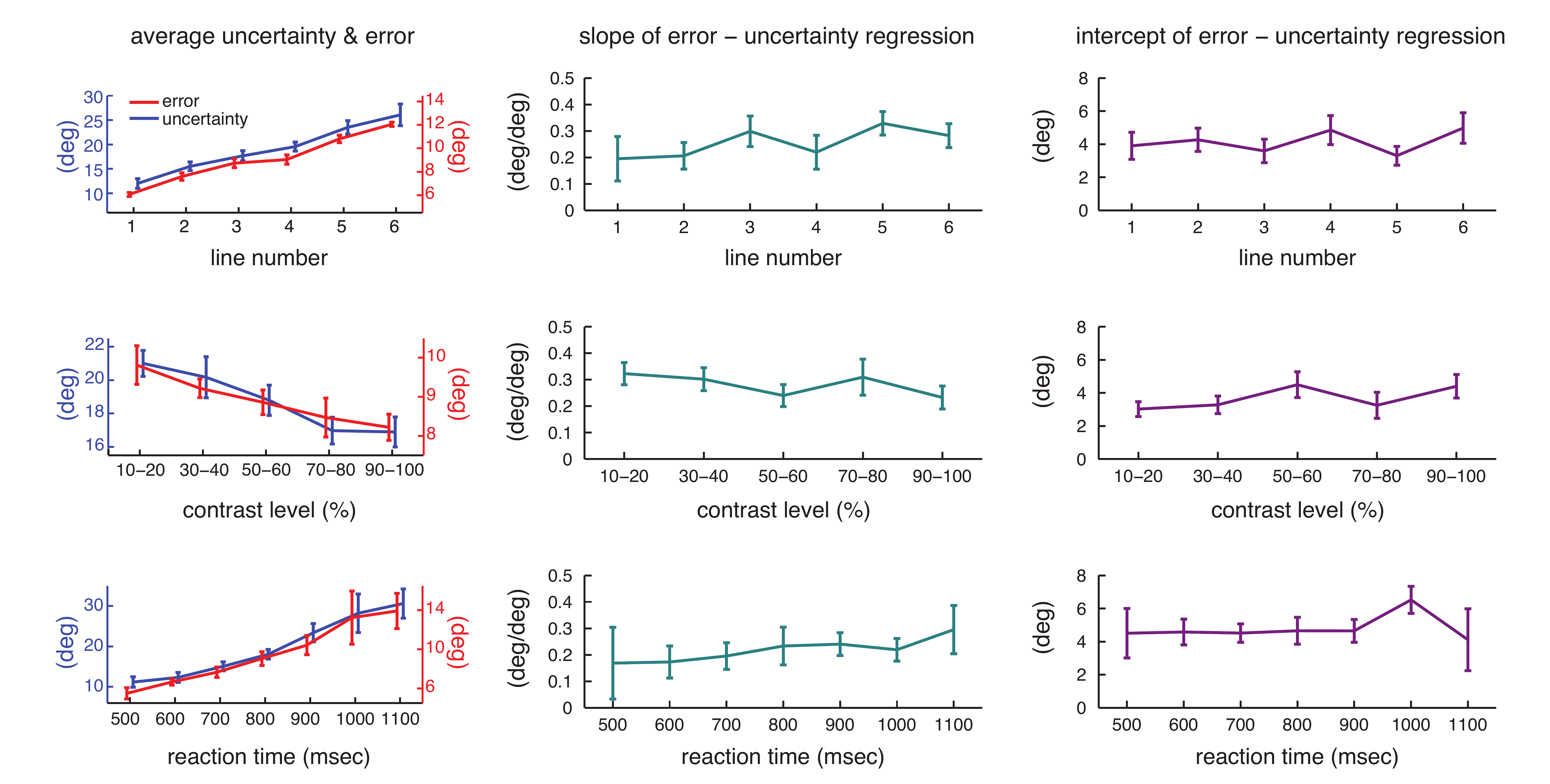}
\caption{\label{fig:errunc_taskdiff}{\bf Average error and uncertainty depend on task difficulty, but not the calibration of uncertainty against error.}
Top row: dependencies on the number of line segments. 
Middle row: dependencies on the contrast level of the target segment. 
Bottom row: dependencies on reaction time.
Left column: average error rates and uncertainty levels.
Middle column: the slope of the error-uncertainty regression.
Right column: the y-intercept of the error-uncertainty regression.}
\end{figure}

In contrast, task difficulty did not have an effect the predictive relationship between error and uncertainty. We quantified the effect of task difficulty on the error-uncertainty regression (gray lines in Fig.~\ref{fig:erruncall}B) by first selecting trials based on either number of line segments, contrast level, or RT duration for each subject, and then taking the slope and intercept of the regression line fitted to data collected in these trials (Figure~\ref{fig:errunc_taskdiff}, right two columns).  Neither the slope nor the intercept were sensitive significantly to changes in any of the three parameters of contrast, line number, or RT (p$>$0.24  for Spearman's correlation in all but one case, for which p$>$0.098, i.e.\ still non-significant).  This means that while task difficulty reliably influenced subjects errors and uncertainty levels, it left the calibration between the two essentially unaltered, suggesting that subjects used a single, universal internal scale of uncertainty -- in line with theories of probabilistic representations.

\subsection{Across-trial correlation between error and uncertainty as a function of stimulus presentation time}

\begin{figure}[h]
\centering \includegraphics[width=0.8\textwidth]{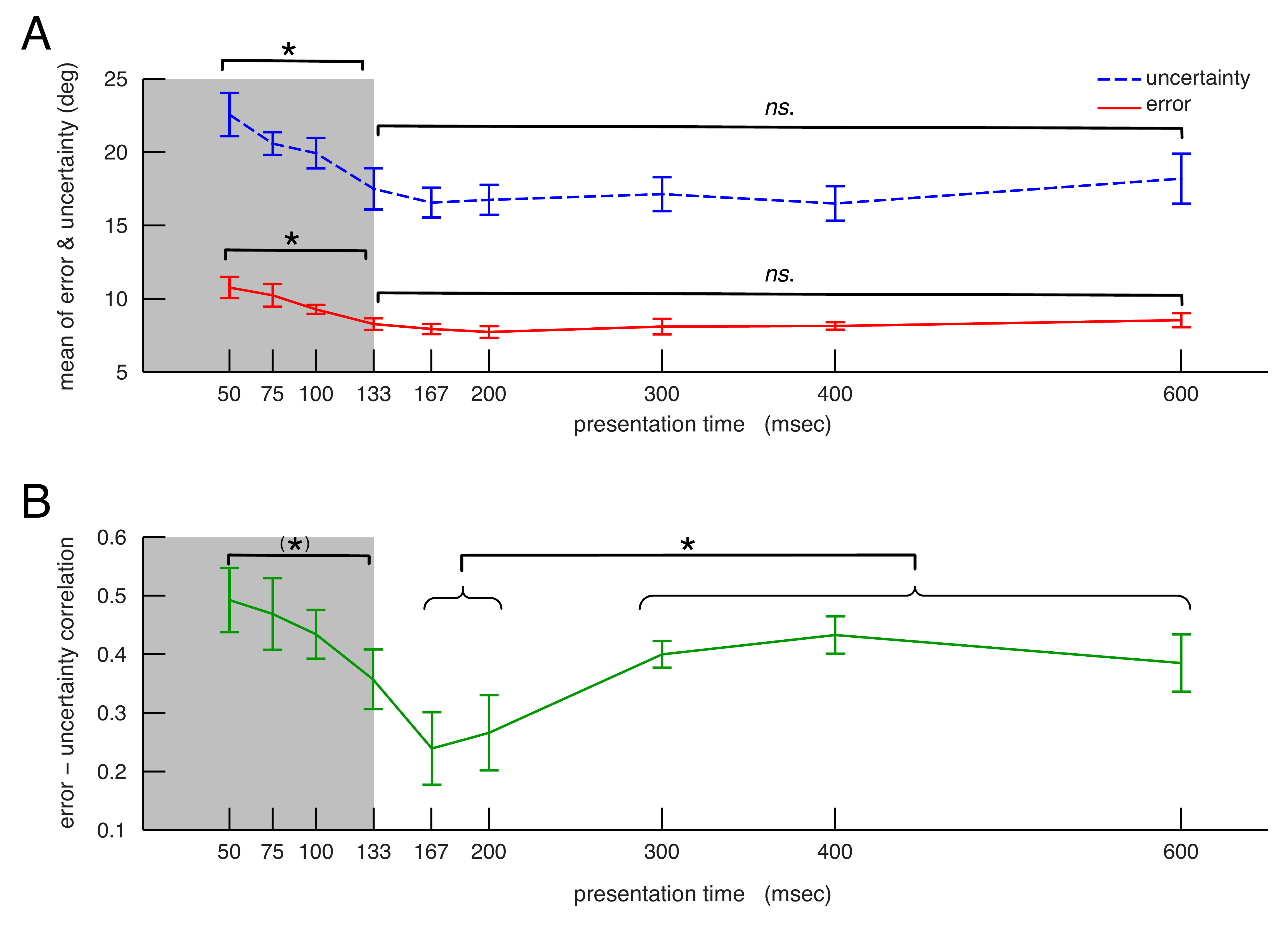}
\caption{\label{fig:errunc_prestime}{\bf Presentation time-dependence of the trial-by-trial correlation between error and uncertainty.}
The rising part of the correlation curve confirms probabilistic sampling during the task.
\textbf{A.} Average error (red) and uncertainty (blue) curves as a function of presentation time. The shaded area represents the cutoff point (obtained independently for error and uncertainty, but coinciding at the same presentation time) above which there was no significant change in values between two neighboring points of the curves. Cutoff points were determined by recursively calculating the significance of the difference between correlations computed from data at a particular presentation time vs.\ data at all longer presentation times. Cutoff points were defined as the first time point (going backwards from the last time point) at which this comparison yielded a significant difference. 
\textbf{B.} Average correlation between error and uncertainty as a function of presentation time. The shaded area is inherited from panel \textbf{A}.  Error bars in both panels represent s.e.m.} 
\end{figure}

To test the main prediction of the present paper, we analyzed error, uncertainty and the correlation between the two as a function of presentation time.  Both subjects' errors and their uncertainty decreased significantly until around 133 msec  (Figure~\ref{fig:errunc_prestime}A, shaded area) and then remained constant for the rest of the duration of the trial (Figure~\ref{fig:errunc_prestime}A, area without shading).   Despite this coordinated decrement in error and uncertainty, the correlation between the two did not increase but instead decreased significantly, as the derivations overviewed in Section 2 predicted it both for EI and PS (Figure~\ref{fig:errunc_prestime}B, shaded area).  However, immediately after error and uncertainty asymptotes in Figure~\ref{fig:errunc_prestime}A, a rapid increase in the correlation began until the curve saturated after 400 msec of presentation time (Figure~\ref{fig:errunc_prestime}B).  A one-way ANOVA including data from beyond the cutoff point showed that the increase in correlation was significant (p$<$ 0.037) and specifically the correlation at time points at the trough of the dip  (167-200 msec) were significantly lower than those at the plateau (300-400-600 msec) of the curve (p$<$0.0014). These results are in close match with the prediction of the PS model and cannot be explained by the EI model.

\section{Discussion}

In this paper, we 
\begin{enumerate}
\item presented a new hypothesis about the role of time in perceptual decision making based on the idea of sequential approximate probabilistic computation in the cortex;
\item derived models and analytical predictions as to how to detect behaviorally whether this proposed role or evidence integration dominates during a decision making task;
\item developed a novel paradigm for testing the models' predictions; and 
\item after conducting the experiment, we confirmed that, indeed, humans show the hallmarks of probabilistic sampling in perceptual decision making tasks.
\end{enumerate}
We envision two aspects of our contribution to be of significance for the field of computational neuroscience.  First, our paradigm involved orientation estimation, arguably one of the simplest perceptual decision making tasks, and our results showed that the uncertainty-error correlation is not affected by dramatic changes in task difficulty, such as 80\% contrast variation or 3-fold increase in the number of potential targets.  This indicates that uncertainty is not an ``add-on'' cognitive metric that is assessed post-perceptually by a separate process, but rather an essential aspect of sensory representation that is inherently encoded together with the feature value of the input from the very beginning of the processing. This result provides a much needed extension of earlier studies proposing approximate probabilistic coding in the cortex by generalizing those claims down to the level of early sensory perception.  

Second, our new paradigm allows us to investigate the effect of sampling-based probabilistic coding with higher sensitivity than it was possible before.  Earlier works explored the issue of sampling with methods that used probability matching as the assay for sampling-based probabilistic coding, and hence they could only hope to detect sampling when it is using one or very few samples  \citep{drugowitsch13, acuna13, vul13}.  This technical difficulty led to some confusions as to whether one can see evidence of sampling in particular tasks. With our method, the effect of sampling can still be tracked in the regime of a few hundred samples.  Indeed, a rough estimate based on our results suggests that in the present task, samples might arrive as fast as one every couple of milliseconds.  This number is feasible only if the cortex uses an overcomplete representation.  Nevertheless, even this speed presents no insurmountable problem to our experimental assay, opening the possibility of conducting experimental tests of probabilistic representations in a wide variety of contexts.

It is important to clarify that even though PS has been introduced in this paper as an alternative to EI, the two are not mutually exclusive and not even clearly separable in time.  Sampling starts from the first moment evidence is obtained, and evidence integration does not ``stop'' at some point, only its benefit vanishes. Thus, an obvious follow-up in our research program is to analyze the relationship between evidence integration and probabilistic sampling in various tasks to shed light on the common and separate aspects of the two processes. Furthermore, although we analyzed PS as a particular approximate inference method the cortex may employ, there could be other such candidates, and it will be interesting to see whether other methods that also iteratively refine the approximate posterior, e.g.\ (loopy) belief propagation or expectation propagation, result in similar temporal trends in the error-uncertainty correlation.

\subsubsection*{Acknowledgments}

We thank useful discussions with R.\ Haefner, M.\ Sahani, P.\ Latham, P.\ Dayan, D.\ Wolpert, G.\ Orb\'an, P.\ Berkes, and J.\ Solomon. This work was supported by an NSF (MP, JF, grant no.\ IOS-1120938), a Marie-Curie grant by the EC (JF, grant no.\ CIG 618918) Wellcome Trust (ML).

\bibliography{lowLevelUncertain}
\bibliographystyle{nature4brief}


\providecommand{\mathsbf}[1]{\text{\textbf{\textsf{#1}}}}

\providecommand{\ttheta}{\boldsymbol{\theta}}
\providecommand{\argmin}{\mathop{\textrm{argmin}}}
\providecommand{\argmax}{\mathop{\textrm{argmax}}}
\providecommand{\dd}{\mathrm{d}}

\providecommand{\fun}[2]{#1\!\left(#2\right)}
\providecommand{\funrm}[2]{\mathrm{#1}\!\left(#2\right)}
\providecommand{\funbf}[2]{\mathbf{#1}\!\left(#2\right)}
\providecommand{\funsbf}[2]{\mathsbf{#1}\!\left(#2\right)}
\providecommand{\functional}[2]{\mathcal{#1}\!\left[\fun{#2}{\cdot}\right]}
\providecommand{\functionalrm}[2]{\mathrm{#1}\!\left[\fun{#2}{\cdot}\right]}

\providecommand{\transpose}[1]{#1^\mathsf{T}}

\providecommand{\norm}[1]{\bigl|#1\bigr|}
\providecommand{\rectify}[1]{\left[#1\right]_+}
\providecommand{\leftlim}[1]{{\displaystyle \mathop{\simeq}_{#1}}}
\providecommand{\rightlim}[1]{{\displaystyle \mathop{\longrightarrow}_{#1}}}

\providecommand{\PP}[1]{\mathrm{P}\!\left(#1\right)}
\providecommand{\PPi}[2]{\mathrm{P}#1\!\left(#2\right)}
\providecommand{\PPestim}[1]{\hat{\mathrm{P}}\!\left(#1\right)}
\providecommand{\PPapprox}[1]{\tilde{\mathrm{P}}\!\left(#1\right)}
\providecommand{\PPapproxi}[2]{\tilde{\mathrm{P}}#1\!\left(#2\right)}
\providecommand{\PPtrue}[1]{\mathrm{P}^*\!\left(#1\right)}
\providecommand{\PPapproxxtrue}[1]{\tilde{\mathrm{P}}^*\!\left(#1\right)}
\providecommand{\PPname}[2]{\mathrm{P_{#1}}\!\left(#2\right)}
\providecommand{\QQ}[1]{\mathrm{Q}\!\left(#1\right)}

\providecommand{\given}{\vert}
\providecommand{\ggiven}{\parallel}

\providecommand{\KL}[2]{\mathrm{KL}\!\left[#1\ggiven #2\right]}
\providecommand{\entropy}[1]{\mathrm{H}\!\left[#1\right]}
\providecommand{\infom}[2]{I\!\left[#1,#2\right]}
\providecommand{\average}[1]{\left< #1 \right>}
\providecommand{\E}[2]{\mathbb{E}_{#1}\!\left[#2\right]}
\providecommand{\Var}[2]{\mathbb{V}_{#1}\!\left[#2\right]}
\providecommand{\Cov}[3]{\mathbb{C}_{#1}\!\left[#2,#3\right]}
\providecommand{\Corr}[3]{\mathrm{Corr}_{#1}\!\left[#2,#3\right]}

\providecommand{\normalpdf}[1]{\fun{\mathcal{N}}{#1}}

\providecommand{\realset}{\mathbb{R}}

\newcommand{\posterior}{\PPname{x}{x\given y}}
\newcommand{\posteriorn}{\PPname{x}{x\given y_1 \ldots y_n}}
\newcommand{\postmean}{\mu_\mathrm{x}}
\newcommand{\postmeanrelkurt}{\postrelkurt}
\newcommand{\poststd}{\sigma_\mathrm{x}}
\newcommand{\postvar}{\sigma^2_\mathrm{x}}
\newcommand{\postvarcubehalf}{\sigma^3_\mathrm{x}}
\newcommand{\postvarsq}{\sigma^4_\mathrm{x}}
\newcommand{\postrelskew}{\xi_\mathrm{x}}
\newcommand{\postrelkurt}{\kappa_\mathrm{x}}

\newcommand{\difficulty}{\gamma}
\newcommand{\difficultysq}{\gamma^2}
\newcommand{\difficultycube}{\gamma^3}
\newcommand{\difficultyquad}{\gamma^4}
\newcommand{\meandifficulty}{\mu_{\difficulty}}
\newcommand{\vardifficulty}{\sigma^2_{\difficulty}}
\newcommand{\meandifficultysq}{\mu_{\difficultysq}}
\newcommand{\vardifficultysq}{\sigma^2_{\difficultysq}}
\newcommand{\relskewdifficulty}{\xi_{\difficulty}}
\newcommand{\relkurtdifficulty}{\kappa_{\difficulty}}
\newcommand{\cvdifficultysq}{\omega_{\difficultysq}}
\newcommand{\cvsqdifficultysq}{\omega^2_{\difficultysq}}
\newcommand{\cvinvsqdifficultysq}{\omega^{-2}_{\difficultysq}}

\newcommand{\difficultynsq}{\gamma_n^2}
\newcommand{\meandifficultynsq}{\mu_{\difficultynsq}}
\newcommand{\vardifficultynsq}{\sigma^2_{\difficultynsq}}
\newcommand{\cvsqdifficultynsq}{\omega^2_{\difficultynsq}}
\newcommand{\postnrelkurt}{\kappa_n}

\newcommand{\samplecorr}{\bar{r}}
\newcommand{\samplecorrN}[1]{\bar{r}_{#1}}
\newcommand{\postmeanest}{\hat{\mu}_\mathrm{x}}
\newcommand{\poststdest}{\hat{\sigma}_\mathrm{x}}
\newcommand{\poststdestt}{\tilde{\sigma}_\mathrm{x}}
\newcommand{\postvarest}{\hat{\sigma}^2_\mathrm{x}}
\newcommand{\postvarsqest}{\hat{\sigma}^4_\mathrm{x}}
\newcommand{\postvarestt}{\tilde{\sigma}^2_\mathrm{x}}
\newcommand{\postvarsqestt}{\tilde{\sigma}^4_\mathrm{x}}

\newcommand{\postmeanrep}{\check{\mu}}
\newcommand{\postvarrep}{\check{\sigma}^2}
\newcommand{\varpostmeanrep}{\epsilon^2_{\postmeanrep}}
\newcommand{\varpostvarrep}{\epsilon^2_{\postvarrep}}
\newcommand{\varsqpostmeanrep}{\epsilon^4_{\postmeanrep}}
\newcommand{\varsqpostvarrep}{\epsilon^4_{\postvarrep}}
\newcommand{\relkurtpostmeanrep}{\kappa_{\postmeanrep}}
\newcommand{\relkurtpostvarrep}{\kappa_{\postvarrep}}

\newcommand{\perf}{\varepsilon}
\newcommand{\corr}{\rho}

\newcommand{\ontop}[2]{\genfrac{}{}{0pt}{1}{#1}{#2}}
\newcommand{\sumi}{\sum_{i=1}^N}
\newcommand{\sumj}{\sum_{\ontop{j=1}{j\neq i}}^N}
\newcommand{\sumJ}{\sum_{j=1}^N}
\newcommand{\sumk}{\sum_{\ontop{k=1}{\ontop{k\neq i}{k\neq j}}}^N}
\newcommand{\sumK}{\sum_{k=1}^N}
\newcommand{\suml}{\sum_{\ontop{l=1}{\ontop{l\neq i}{\ontop{l\neq j}{l\neq k}}}}^N}
\newcommand{\sumL}{\sum_{l=1}^N}

\renewcommand{\thefootnote}{\roman{footnote}}


\newpage

\setcounter{section}{0}
\setcounter{equation}{0}

\begin{center}
{\Large \textbf{Supplementary Information}}
\end{center}

\section{Introduction}

We consider an estimation task, in which subjects have to estimate an environmental variable, $x\in\realset$ (which we assume to be scalar for simplicity). In each trial, the subject receives partial information about the true value of $x$, $x^*$, through making observation $y$ and is asked to report two things: 
\begin{enumerate}
\item their best estimate of $x^*$  -- assuming that they optimise a squared error penalty between this estimate and $x^*$;
\item their subjective uncertainty about this estimate, which we take to be their expected squared error between their estimate and $x^*$.
\end{enumerate}
We will be interested in how subjective uncertainty, the actual squared error of the estimate, and the correlation between these two quantities change as a function of time (within in a trial) -- where time is going to parametrise some key aspects of the model.

\section{Exact representation of uncertainty}

A complete representation of uncertainty means that the subject represents $\posterior$, a posterior distribution over $x$ given the partial information available to them in $y$. First, in this section, we consider the case when the amount of information received about $x$ does not change over the trial, ie.\ $\posterior$ remains fixed, and the representation of uncertainty is \emph{exact}, i.e.\ the relevant parameters of $\posterior$ are directly and exactly accessible by the subject. This will then serve as the starting point for derivations regarding evidence integration (EI, sections \ref{sec:EI} and \ref{sec:behnoise}) and probabilistic sampling (PS, sections \ref{sec:approxrepr}-\ref{sec:PS}). 

We introduce the following quantities to characterise the statistics of $x$ under $\posterior$:
\begin{align}
\postmean &= \E{x}{x} && \text{mean}\\
\postvar &= \E{x}{\left(x-\postmean\right)^2} && \text{variance} \\
\postrelskew &= \E{x}{\left(x-\postmean\right)^3} / \postvarcubehalf && \text{skewness}\\
\postrelkurt &= \E{x}{\left(x-\postmean\right)^4} / \postvarsq -3 && \text{excess kurtosis} 
\end{align}
(As an example, $\postrelskew=\postrelkurt=0$ when $\posterior$ is a normal distribution.)

Note that, in general, $\posterior$ is going to be different on each trial using the same $x^*$ because the particular information given by the observation $y$ is different. Thus, the moments of $\posterior$ will also be different on each trial, that is, importantly, they are random variables themselves. For simplicity, we will assume in the following that while $\postmean$ and $\postvar$ change across trials, $\postrelskew$ and $\postrelkurt$ remain constant -- i.e. the posterior only changes in its location and scale but not in its shape otherwise.

\subsection{Task difficulty, $\difficulty$, and properties of the posterior mean, $\postmean$, and variance $\postvar$}

An exact (and consistent) representation of uncertainty means that the true value of $x$, $x^*$, behaves as if it was actually sampled from $\posterior$. To simplify notation, however, and without loss of generality, we will assume that on each trial $x^*=0$ and shift the reference frame for $x$ accordingly. 

In order to minimise squared error, the best estimate of $x^*$ the subject can report is their posterior mean, $\postmean$. Remember (see above), that $\postmean$ is itself a random variable. In fact, given our shift of reference frame, the distribution of $\postmean$ around $x^*=0$ is the mirror image of the distribution of $x$ around $\postmean$, ie.\ $\PP{\postmean=x'}=\PPname{x}{\postmean-x'}$. 

We define the variance of $\postmean$ as the (squared) difficulty of a trial, $\difficulty$. This implies that the mean, variance, and fourth (central) moment of $\postmean$ are
\begin{align}
\E{\postmean\given\difficulty}{\postmean}&=x^*=0 \\
\E{\postmean\given\difficulty}{\postmean^2}&=\difficultysq \\
\E{\postmean\given\difficulty}{\postmean^4}&=\left(\postmeanrelkurt+3\right)\,\difficultyquad
\end{align}

where we have made use of the fact that the distribution of $\postmean$ is the `mirror image' of the posterior (see above) and that it is centred on $x^*=0$ and thus its variance is equal its mean squared error. 

Note that, in general, $\difficulty$ itself might also change from trial-to-trial, as some trials might be more or less difficult, or we may be paying more or less attention to the stimuli, so we will treat it also as a random variable with the following statistics:
\begin{align}
\E{\difficulty}{\difficulty}&=\meandifficulty\\
\Var{\difficulty}{\difficulty}&=\vardifficulty\\
\E{\difficulty}{\difficultysq}&=\meandifficultysq=\meandifficulty^2+\vardifficulty\\
\Var{\difficulty}{\difficultysq}&=\vardifficultysq 
\end{align} 
It will also be useful later to define the squared coefficient of variation of $\difficultysq$: $\cvsqdifficultysq=\vardifficultysq/\meandifficultysq^2\geq0$.

A consistent representation of subjective uncertainty should be predictive of the errors the subject makes. Ideally thus, in the above estimation task, the reported uncertainty should be $\postvar$, the variance of the posterior, which will be equal to the variance of $\postmean$, which in turn by definition is $\difficultysq$. 
\begin{equation}
\postvar\given\difficulty=\difficultysq
\end{equation}
which in turn entails that the following is true for any $k$:
\begin{align}
\E{\poststd\given\difficulty}{\poststd^k}&=\difficulty^k
\end{align}
It is also useful to note that the same equality (ie.\ that the distribution of $\postvar$ given $\difficulty$ is a delta distribution) also implies that $\postvar$ and $\postmean$ are independent given $\difficulty$.

\subsection{Some useful identities \label{sec:useful identities}}

We use the formula $\E{X}{\fun{f}{X}}=\E{Y}{\E{X\given Y}{\fun{f}{X}}}$ to derive the following.

\begin{align}
\E{\postmean}{\postmean} &= \E{\difficulty}{\E{\postmean\given\difficulty}{\postmean}} &&= \E{\difficulty}{0}  &&= 0 \\
\E{\postmean}{\postmean^2} &= \E{\difficulty}{\E{\postmean\given\difficulty}{\postmean^2}} &&= \E{\difficulty}{\difficultysq} &&= \meandifficultysq \\
\E{\postmean}{\postmean^4} &= \E{\difficulty}{\E{\postmean\given\difficulty}{\postmean^4}} &&= \E{\difficulty}{\left(\postmeanrelkurt+3\right)\,\difficultyquad} &&=  \left(\postmeanrelkurt+3\right)\, \left(\meandifficultysq^2 + \vardifficultysq \right)\\
\E{\poststd}{\postvar} &= \E{\difficulty}{\E{\poststd\given\difficulty}{\postvar}} &&= \E{\difficulty}{\difficultysq}  &&= \meandifficultysq \\
\E{\poststd}{\postvarsq} &= \E{\difficulty}{\E{\poststd\given\difficulty}{\postvarsq}} &&= \E{\difficulty}{\difficultyquad}  &&= \meandifficultysq^2 + \vardifficultysq\\
\E{\postmean,\poststd}{\postmean^2\,\postvar}  &= \E{\difficulty}{\E{\postmean,\poststd \given \difficulty}{\postmean^2\,\postvar}} = \E{\difficulty}{\E{\postmean \given \difficulty}{\postmean^2}\,\E{\poststd \given \difficulty}{\postvar}} &&= \E{\difficulty}{\difficultysq\,\difficultysq} &&= \meandifficultysq^2 + \vardifficultysq\\
\E{\postmean,\poststd}{\postmean\,\postvarcubehalf}  &= \E{\difficulty}{\E{\postmean,\poststd \given \difficulty}{\postmean\,\postvarcubehalf}} = \E{\difficulty}{\E{\postmean \given \difficulty}{\postmean}\,\E{\poststd \given \difficulty}{\postvarcubehalf}} &&= \E{\difficulty}{0\,\difficultycube} &&= 0
\end{align}

\subsection{Performance}

The average squared error for an exact representation of uncertainty is simply the squared error of the mean that is
\begin{equation}
\perf=\E{\postmean}{\postmean^2} =  \meandifficultysq
\end{equation}

\subsection{Correlation between error and uncertainty}

Now, we can ask the question how much the error in the estimate is going to correlate with the subjective representation of uncertainty. For this we first compute the covariance of these two quantities:

\begin{align}
\Cov{\postmean,\poststd}{\postmean^2}{\postvar}
&= \E{\postmean,\poststd}{\postmean^2 \, \postvar}-\E{\postmean}{\postmean^2}\,\E{\poststd}{\postvar}
&&= \meandifficultysq^2 + \vardifficultysq - \meandifficultysq \, \meandifficultysq 
&&= \vardifficultysq\\
\end{align}
Likewise, we can compute the respective variances:
\begin{align}
\Var{\postmean}{\postmean^2}
&=\E{\postmean}{\postmean^4}-\left(\E{\postmean}{\postmean^2}\right)^2
&&= \left(\postmeanrelkurt+3\right)\, \left(\meandifficultysq^2 + \vardifficultysq \right) - \meandifficultysq^2
&&= \left(\postmeanrelkurt+3\right)\, \vardifficultysq  +  \left(\postmeanrelkurt+2\right)\, \meandifficultysq^2 \\
\Var{\poststd}{\postvar}
&=\E{\poststd}{\postvarsq}-\left(\E{\poststd}{\postvar}\right)^2
&&=\meandifficultysq^2 + \vardifficultysq - \meandifficultysq^2
&&=\vardifficultysq
\end{align}

So the correlation between squared error and subjective uncertainty is
\begin{align}
\corr
& =\frac{\vardifficultysq}{\sqrt{\left(\left(\postmeanrelkurt+3\right)\vardifficultysq+\left(\postmeanrelkurt+2\right)\meandifficultysq^2\right)\,\vardifficultysq}}\\
\corr&=\frac{1}{\sqrt{\postmeanrelkurt+3+\left(\postmeanrelkurt+2\right)\,\cvinvsqdifficultysq}} \label{eq:corr_exact}
\end{align}

There are three things to note about this result:
\begin{enumerate}
\item The correlation is \emph{not} going to be $1$ in the general case, even though we are talking about an \emph{exact} representation of uncertainty. How can that be? This is  because only the expected value of the squared error, not the squared error itself is predicted by uncertainty, otherwise we could be cheating (i.e.\ if we could predict the error itself, than we might as well subtract it from our estimate thus making our error zero).\footnote{Well, we wouldn't quite be cheating in a special case: when the error is known only up to its sign. In that case, the squared error is known exactly but we still can't subtract away our error because we don't know its sign. Indeed, in that case the distribution of $\postmean$ is a mixture of two delta distributions (with equal weights) sitting at equal distances from zero, and thus $\postmeanrelkurt=-2$, in which case the formula in Eq.~\ref{eq:corr_exact} correctly predicts $\corr=1$.}
\item The correlation is going to be zero when $\cvsqdifficultysq\rightarrow 0$, so we need trials with varying difficulty to get meaningful results. (This is was expected intuitively, as $\cvsqdifficultysq\rightarrow 0$ means that all trials have the same difficulty, and associated posterior variance, and therefore the same uncertainty estimate -- and one obtains a trivial zero correlation between two quantities one of which does not ever change.)
\item Nothing changes with time here -- simply because we ascribed no role for time in the model.
\end{enumerate}

\section{Evidence integration \label{sec:EI}}

We now consider the case of a probabilistic representation that is still exact, as before, but it is representing a posterior distribution that is changing in time because it is conditioned on increasing amounts of data, $\posteriorn$. 

In general, the variance of such a posterior scales inversely with $n$, the number of data points. Using the equivalence of difficulty, $\difficultysq$ and posterior variance $\postvar$ described in the previous section, this means that the ``effective difficulty'' of trials in which $n$ data points have been observed can be expressed as 
\begin{equation}
\difficultynsq=\difficultysq/n
\end{equation}
where we use $\difficultysq$ here to describe the difficulty of trials with a single data point. In turn, this means that the relevant moments characterising the distribution of $\difficultynsq$ scale as
\begin{align}
\meandifficultynsq&=\meandifficultysq/n\\
\vardifficultynsq&=\vardifficultysq/n^2
\end{align}
Note that this scaling implies that the coefficient of variation, which ultimately matters for the analysis of error-variance correlations remains constant, and so
\begin{equation}
\cvsqdifficultynsq=\cvsqdifficultysq
\end{equation}

Beside the variance, the (excess) kurtosis of $\posteriorn$ is also expected to change and converge towards zero, as the posterior tends to converge to a normal distribution with increasing amounts of data (modulo degenaracies) -- this is the argument on which the usual Laplace approximation is based. We will make the assumption\footnote{See Appendix.} that the kurtosis of the distribution also scales as $1/n$
\begin{equation}
\postnrelkurt=\postrelkurt/n
\end{equation}
where we (re-)use $\postrelkurt$ to denote the kurtosis of the posterior after observing a single data point.

Putting the above scalings together and substituting them to the results of the previous section, we can now derive the evolution of estimation errors and error-uncertainty correlations in the context of evidence integration  as the following:
\begin{align}
\perf &= \frac{\meandifficultysq}{n}\\
\corr
&=\frac{1}{\sqrt{\postmeanrelkurt/n+3+\left(\postmeanrelkurt/n+2\right)\,\cvinvsqdifficultysq}}
\end{align}

\section{Approximate representations of uncertainty \label{sec:approxrepr}}

We now return to the case of a static posterior distribution. In an approximate representation of uncertainty, the posterior mean and variance are not directly accessible, only their approximate estimates, $\postmeanest$ and $\postvarest$, respectively, which can be noisy and / or biased. 

In this case, the performance and correlation between the reported uncertainty and actual squared error can be computed as
\begin{align}
\perf&=\E{\postmeanest}{\postmeanest^2} &&= \E{\postmean,\poststd}{\E{\postmeanest\given\postmean,\poststd}{\postmeanest^2}}\\
\corr&=\frac{\Cov{\postmeanest,\poststdest}{\postmeanest^2}{\postvarest}}{\sqrt{\Var{\postmeanest}{\postmeanest^2}\,\Var{\poststdest}{\postvarest}}}\\
\intertext{where}
\Var{\postmeanest}{\postmeanest^2}&=\E{\postmeanest}{\postmeanest^4}-\left(\E{\postmeanest}{\postmeanest^2}\right)^2 &&=\E{\postmean,\poststd}{\E{\postmeanest\given\postmean,\poststd}{\postmeanest^4}}-\left(\E{\postmean,\poststd}{\E{\postmeanest\given\postmean,\poststd}{\postmeanest^2}}\right)^2\\
\Var{\poststdest}{\postvarest}&=\E{\poststdest}{\postvarsqest}-\left(\E{\poststdest}{\postvarest}\right)^2 &&=\E{\postmean,\poststd}{\E{\poststdest\given\postmean,\poststd}{\postvarsqest}}-\left(\E{\postmean,\poststd}{\E{\poststdest\given\postmean,\poststd}{\postvarest}}\right)^2\\
\Cov{\postmeanest,\poststdest}{\postmeanest^2}{\postvarest}&=\E{\postmeanest,\poststdest}{\postmeanest^2\,\postvarest}-\E{\postmeanest}{\postmeanest^2}\,\E{\poststdest}{\postvarest}&&=\E{\postmean,\poststd}{\E{\postmeanest,\poststdest\given\postmean,\poststd}{\postmeanest^2\,\postvarest}}-\nonumber\\
&&&\phantom{=}-\E{\postmean,\poststd}{\E{\postmeanest\given\postmean,\poststd}{\postmeanest^2}}\,\E{\postmean,\poststd}{\E{\poststdest\given\postmean,\poststd}{\postvarest}}
\end{align}
Therefore, in order to be able to compute these quantities of interest, we need to know the following five statistics of the approximate representation:
\begin{align}
\E{\postmeanest\given\postmean,\poststd}{\postmeanest^2} && \E{\postmeanest\given\postmean,\poststd}{\postmeanest^4}\\
\E{\poststdest\given\postmean,\poststd}{\postvarest} && \E{\poststdest\given\postmean,\poststd}{\postvarsqest}\\
&& \E{\postmeanest,\poststdest\given\postmean,\poststd}{\postmeanest^2\,\postvarest}
\end{align}

A subtle note about approximate representations: remember, that the posterior variance was only relevant for the subject in the context of the estimation task inasmuch as it predicted the average squared error between the posterior mean and the true stimulus value -- which was the case for an exact representation of uncertainty. However, for an approximate representation of uncertainty, the total estimation error is compounded by the fact that the posterior mean itself is also only estimated. Thus, ideally, the subject should estimate their own estimation error of the posterior mean, based on their estimate of the posterior mean and variance, and add it to the estimate of the posterior variance, and report this total error estimate, $\postvarestt$, as their uncertainty. We will return to this point later in the context of sampling-based representations in section \ref{sec:impunc}.

\section{Sampling-based representations \label{sec:PS}} 

We now consider the case of sampling-based representations, that is when we do not have direct access to the relevant parameters of the posterior, $\posterior$, such as $\postmean$ and $\postvar$, but only to a finite number, $N$, of samples $x_i,\,i=1\ldots N$, that are iid.\ according to $\posterior$. We will start by considering some statistical properties of these samples that will be useful for the derivations later. We will then derive unbiased estimators that estimate the mean and variance of the posterior from the samples. Next, as a side, we will analyse the basic statistical properties of these estimators, namely their variances and covariance (and correlations). Then, we will derive the correct measure of uncertainty in the context of a sampling based estimator, that takes into account its own estimation errors, and derive the statistical properties of the mean and uncertainty estimate that are directly relevant for computing the performance and error-variance correlation. Finally, by combining results from previous sections we will derive how performance and error-variance changes with the number of samples.

\subsection{Statistics of the samples}

The fact that $x_i,\,i=1\ldots N$ are identically distributed according to $\posterior$ of course entails that the mean and variance (and other moments) of any individual sample will match the corresponding moments of the posterior such that
\begin{align}
\E{x}{x_i}&=\postmean\\
\Var{x}{x_i}&=\postvar\\
\E{x}{\left(x_i-\postmean\right)^3} &=\E{x}{\left(x-\postmean\right)^3}=\postrelskew \, \postvarcubehalf\\
\E{x}{\left(x_i-\postmean\right)^4} &=\E{x}{\left(x-\postmean\right)^4}=\left(\postrelkurt + 3\right) \, \postvarsq
\end{align}

\subsection{Unbiased estimators of the mean and variance of the posterior}

We will use standard unbiased estimators (and their extensions) for estimating the posterior mean and variance from the iid.\ samples. (As a consequence, the squared error of these estimators will be their variance, so we will use these terms interchangeably in this context.) From here on, to simplify the algebra, we will assume (without loss of generality) that $\postmean=0$ -- unless otherwise noted. (Note that this does not imply $\postmeanest=0$.)

The unbiased estimator for the mean of the posterior is simply the sample mean:
\begin{align}
\postmeanest&=\frac{1}{N} \, \sum_{i=1}^N x_i
\end{align}
and the unbiased estimator for the variance is the sample variance:
\begin{align}
\postvarest&=\frac{1}{ N - 1} \, \sum_{i=1}^N \left(x_i - \postmeanest\right)^2 = \frac{1}{N - 1}  \, \left( \sum_{i=1}^N x_i^2 - N \, \postmeanest^2\right) 
\end{align}

\subsection{Basic statistics of the sample mean and variance}

\subsubsection{Variance of the sample mean}

\begin{align}
\Var{x}{\postmeanest}
&= \E{x}{\postmeanest^2}-\left(\E{x}{\postmeanest}\right)^2\\
&= \frac{1}{N^2} \sum_i \sum_j \E{x}{x_i \, x_j}-\postmean^2\\
&= \frac{1}{N^2} \left(N \, \Var{x}{x_i} + \sum_{i=1}^N \, \sum_{j=1, j\neq i}^N \Cov{x}{x_i}{x_j} \right)\\
&= \frac{1}{N}  \, \postvar
\end{align}

\subsubsection{Variance of the sample variance}

In the following, we will make use of the following identities -- assuming $\postmean=0$:
\begin{align}
\sumi \, \sumJ \E{x}{x_i^2 \, x_j^2}
& = N \, \left(\postrelkurt + 3\right) \, \postvarsq + N \left(N-1\right) \, \postvarsq\\
\sumi \, \sumJ \, \sumK \E{x}{x_i^2 \, x_j \, x_k} 
&= N \,  \left(\postrelkurt + 3\right) \, \postvarsq +  N \left(N-1\right) \, \postvarsq\\
\sumi \, \sumJ \, \sumK \, \sumL \E{x}{x_i \, x_j \, x_k  \, x_l} 
&= N \,  \left(\postrelkurt + 3\right) \, \postvarsq + 3N \left(N-1\right) \, \postvarsq
\end{align}

And so the variance of the sample variance is (omitting several lines of tedious but trivial algebra)
\begin{align}
\Var{x}{\postvarest}
&= \E{x}{\postvarsqest}-\left(\E{x}{\postvarest}\right)^2\\
&= \left(\frac{\postrelkurt}{N} + \frac{2}{N - 1} \right) \, \postvarsq
\end{align}

\subsubsection{Covariance between the sample mean and the sample variance}

In the following, we will make use of the following identities:
\begin{align}
\sumi \, \sumJ \E{x}{x_i^2 \, x_j}
&= N \, \postrelskew  \, \postvarcubehalf \\
\sumi \, \sumJ \, \sumK  \E{x}{x_i \, x_j \, x_k } 
&= N \, \postrelskew  \, \postvarcubehalf
\end{align}

So the covariance between the sample mean and sample variance is (again omitting several lines of tedious but trivial algebra)
\begin{align}
\Cov{x}{\postmeanest}{\postvarest} 
&= \E{x}{\postmeanest\,\postvarest}-\E{x}{\postmeanest}\,\E{x}{\postvarest}\\
&= \frac{\postrelskew}{N} \, \postvarcubehalf
\end{align}
Note that this covariance is $0$ if the posterior is normal, or in general when it is symmetric (because $\postrelskew=0$).

Although we don't need this in the following, but it may be useful to note that the correlation between the sample mean and variance is thus
\begin{align}
\Corr{x}{\postmeanest}{\postvarest} 
&=\frac{\frac{\postrelskew}{N} \, \postvarcubehalf}{\sqrt{\frac{1}{N}  \, \postvar  \, \left(\frac{\postrelkurt}{N} + \frac{2}{N - 1} \right) \, \postvarsq}}\\
&=\frac{\postrelskew}{\sqrt{\postrelkurt + \frac{2N}{N - 1}}}
\end{align}

\subsection{An improved estimator of uncertainty}\label{sec:impunc}

Remember that the objective of a representation of uncertainty is that it should be predictive of the squared error of one's estimate. In a sampling based representation the best estimator (in a squared loss sense) is the sample mean, $\postmean$. The error of this estimator is
\begin{align}
\E{x,\postmean}{\postmeanest^2} &=\E{\postmean}{\postmean^2} + \frac{1}{N} \, \postvar = \left(1 + \frac{1}{N}\right) \, \postvar 
\end{align}
Of course, we don't have direct access to $\postvar$, but we have an unbiased estimator for it, $\postvarest$, using which we can derive the improved estimator of uncertainty as 
\begin{equation}
\postvarestt=\left(1 + \frac{1}{N}\right) \, \postvarest 
\end{equation}

\subsection{Statistical properties relevant for performance and correlation analysis}

\subsubsection{Statistics of the sample mean}

In this section, we will make use of the following identities -- without assuming $\postmean=0$:
\begin{align}
\E{x}{x_i}&=\postmean\\
\E{x}{x_i^2}&=\postvar+\postmean^2\\
\E{x}{x_i^3}&=\postrelskew\postvarcubehalf+3\postmean\postvar+\postmean^3\\
\E{x}{x_i^4}&=\left(\postrelkurt+3\right)\postvarsq+4\postrelskew\postmean\postvarcubehalf+6\postmean^2\postvar+\postmean^4
\end{align}

Now we can derive the relevant statistics of the sample mean (once again omitting several lines of tedious but trivial algebra):
\begin{align}
\E{x}{\postmeanest^2}
&=\frac{1}{N^2}  \sumi \, \sumJ \E{x}{x_i \, x_j} \label{eq:Epostmeansq_sample_a}\\
& =\postmean^2 + \frac{1}{N} \,\postvar \label{eq:Epostmeansq_sample_b} \\
\end{align}
and
\begin{align}
\E{x}{\postmeanest^4}
&=\frac{1}{N^4} \sumi \, \sumJ \, \sumK \, \sumL \E{x}{x_i \, x_j \, x_k  \, x_l} \label{eq:Epostmeanquad_sample_a} \\
&= \postmean^4+ \frac{6}{N} \, \postmean^2\postvar + \frac{4}{N^2} \, \postrelskew\postmean\postvarcubehalf + \frac{\postrelkurt+3N}{N^3}\,\postvarsq \label{eq:Epostmeanquad_sample_b}
\end{align}

\subsubsection{Statistics of the improved uncertainty estimator}

\begin{align}
\E{x}{\postvarestt}&=\left(1 + \frac{1}{N}\right) \, \E{x}{\postvarest} && = \left(1 + \frac{1}{N}\right) \, \postvar \label{eq:Epostvar_sample}\\
\E{x}{\postvarsqestt}&= \left(1 + \frac{1}{N}\right)^2 \, \E{x}{\postvarsqest} &&= \left(1 + \frac{1}{N}\right)^2 \, \left(1+\frac{\postrelkurt}{N} + \frac{2}{N - 1}\right) \, \postvarsq \label{eq:Epostvarsq_sample}
\end{align}

\subsubsection{Joint statistics of the sample mean and the improved uncertainty estimator}

(We once again omit several lines of tedious but trivial algebra.)
\begin{align}
\E{x}{\postmeanest^2 \, \postvarestt}
&=\left(1 + \frac{1}{N}\right) \, \E{x}{\postmeanest^2 \, \postvarest} \label{eq:Epostmeansqpostvar_sample_a} \\
&=\left(1 + \frac{1}{N}\right) \, \frac{1}{N} \, \left(
\left( \frac{\postrelkurt}{N}+1\right) \, \postvarsq+ 
2 \, \postrelskew\,\postmean\,\postvarcubehalf + N\,\postmean^2\,\postvar \right) \label{eq:Epostmeansqpostvar_sample_b}
\end{align}

\subsection{Performance and error-uncertainty correlation with a sampling-based representation}

Note that the quantities 
\begin{align*}
\E{x}{\postmeanest^2} &\text{(Eqs.~\ref{eq:Epostmeansq_sample_a}-\ref{eq:Epostmeansq_sample_b})}& \E{x}{\postmeanest^4} & \text{(Eqs.~\ref{eq:Epostmeanquad_sample_a}-\ref{eq:Epostmeanquad_sample_b})} \\
\E{x}{\postvarestt} & \text{(Eq.~\ref{eq:Epostvar_sample})} & \E{x}{\postvarsqestt} & \text{(Eq.~\ref{eq:Epostvarsq_sample})} \\
&& \E{x}{\postmeanest^2\,\postvarestt} & \text{(Eqs.~\ref{eq:Epostmeansqpostvar_sample_a}-\ref{eq:Epostmeansqpostvar_sample_b})}
\end{align*}
derived in the previous section are exactly the quantities (using slightly different notation as required by the different contexts)
\begin{align*}
\E{\postmeanest\given\postmean,\poststd}{\postmeanest^2} && \E{\postmeanest\given\postmean,\poststd}{\postmeanest^4}\\
\E{\poststdestt\given\postmean,\poststd}{\postvarestt} && \E{\poststdestt\given\postmean,\poststd}{\postvarsqestt}\\
&& \E{\postmeanest,\poststdestt\given\postmean,\poststd}{\postmeanest^2\,\postvarestt}
\end{align*}
that are required to compute the error and error-variance correlation of sampling as an approximate representation in section \ref{sec:approxrepr}.

So, by making the appropriate substitutions (also from section \ref{sec:useful identities}), we can derive the mean squared error of a sampling based representation as 
\begin{align}
\perf
&= \E{\postmeanest}{\postmeanest^2} \\
&=\left(1+\frac{1}{N}\right) \, \meandifficultysq
\end{align}

Similarly, the variance of the error and the uncertainty are:
\begin{align}
\Var{\postmeanest}{\postmeanest^2}
&= \E{\postmean,\poststd}{\E{\postmeanest\given\postmean,\poststd}{\postmeanest^4}}-\left(\E{\postmean,\poststd}{\E{\postmeanest\given\postmean,\poststd}{\postmeanest^2}}\right)^2 \\
&=\left(1 + \frac{1}{N}\right) \, \left(\vardifficultysq + \frac{1}{N} \vardifficultysq + \frac{\postrelkurt}{N^2} \,  \left(\meandifficultysq^2 + \vardifficultysq \right) \right)\end{align}

And so, finally, the error-variance correlation is
\begin{align}
\hspace*{-2cm}\corr
&= \frac{1 + \frac{1}{N}  + \frac{\postrelkurt}{N^2} \,  \left(\cvinvsqdifficultysq + 1 \right)}
{\sqrt{\left(\postmeanrelkurt+3  + \left(\postmeanrelkurt+2\right) \, \cvinvsqdifficultysq  + \frac{1}{N} \left(6 +4 \, \cvinvsqdifficultysq \right)  + \frac{1}{N^2} \left(3  + 2 \, \cvinvsqdifficultysq \right)+ \frac{\postrelkurt}{N^3}\, \left(\cvinvsqdifficultysq + 1 \right)\right) \, 
\left(1 + \left(\frac{\postrelkurt}{N} + \frac{2}{N - 1}\right) \,  \left( \cvinvsqdifficultysq + 1 \right) \right)}}
\end{align}

The following are worth noting about these results:
\begin{enumerate}
\item The squared error, $\perf$, decays as $1/N$ to the squared error of an exact representation of uncertainty, $\meandifficultysq$.
\item The behaviour of the correlation $\corr$ is more complicated and depends on two quantities: $\cvdifficultysq$, quantifying how variable is the difficulty of trials across the experiment, and $\postrelkurt$, the (excess) kurtosis of the posterior.
\end{enumerate}

As a brief summary of the time-dependence of $\corr$, we can note the following:
\begin{enumerate}

\item The correlation, $\corr$ is \emph{not} always monotonically increasing as a function of the number of samples, $N$. In particular, it can be a (transiently) decreasing function of $N$ when the following two conditions are satisfied:
\begin{enumerate}
\item $N$ is small,
\item the kurtosis of the posterior, $\postrelkurt$ is greater than zero.
\end{enumerate}

\item Importantly, the transient decrease in $\corr$ is shorter for higher values of $\cvdifficultysq$ (darker blues), that is, if trials are sufficiently variable in terms of their difficulty, $\cvdifficultysq>1$. In this case, the transient is over by the time performance, measured by $\perf$, converges.

\item After this transient, $\corr$ is indeed a monotonically increasing function of $N$ and converges from below to the correlation exhibited by an exact representation of uncertainty.

\item The time constant of the convergence of $\corr$ can be substantially longer than the time constant of the convergence of $\perf$, in particular for higher values of $\postrelkurt$ and lower values of $\cvdifficultysq$. This means that the in general, $\corr$ will continue to increase when $\perf$ is not discernibly changing anymore.
\end{enumerate}

Note that each of these features seem to be borne out in the data:
\begin{enumerate}

\item There is a transient decrease in $\corr$.

\item This transient is over by the time $\perf$ stabilises.

\item After this transient, $\corr$ increases.

\item This increase continues beyond the time when $\perf$ stabilises.

\end{enumerate}

\section{Behavioural noise \label{sec:behnoise}}

One more important factor may influence the results derived so far: behavioural noise may corrupt the reported posterior mean, $\postmeanrep$, and variance, $\postvarrep$. This is particularly relevant for evidence integration, where error is predicted to asymptote at zero (see above) which is clearly unrealistic. As we will see below, behavioural noise fixes this problem by yielding an above-zero asymptote for error.

We will assume that the behavioural noise in reporting these quantities is unbiased and independent (from each other as well as from $\difficulty$) and has variance $\varpostmeanrep$ and $\varpostvarrep$, respectively, and derive these effects for an exact representation (useful for the static case as well as for the case of evidence integration), and for a sampling-based representation.

In this case, the relevant quantities are
\begin{align}
\Var{}{\postmeanrep}&=\Var{}{\postmeanest}+\varpostmeanrep \\
\Var{}{\postmeanrep^2}&=\Var{}{\postmeanest^2}+\left(\relkurtpostmeanrep+2\right)\,\varsqpostmeanrep +4\,\varpostmeanrep\,\Var{}{\postmeanest}\\
\Var{}{\postvarrep}&=\Var{}{\postvarest}+\varpostvarrep \\
\Cov{}{\postmeanrep^2}{\postvarrep}&=\Cov{}{\postmeanest^2}{\postvarest} \\
\end{align}

\subsection{Exact representation}

The argument above gives the following formul\ae{} for the squared error and error-variance correlation in the static case:
\begin{align}
\perf&=\meandifficultysq+\varpostmeanrep\\
\corr&=\frac{\vardifficultysq}{\sqrt{\left(\left(\postmeanrelkurt+3\right)\vardifficultysq+\left(\postmeanrelkurt+2\right)\meandifficultysq^2 + \left(\relkurtpostmeanrep+2\right)\,\varsqpostmeanrep +4\,\varpostmeanrep\,\meandifficultysq\right)\,\left(\vardifficultysq+\varpostvarrep\right) }}\\
&=\frac{1}{\sqrt{\left(\postmeanrelkurt+3+\left(\postmeanrelkurt+2\right)\cvinvsqdifficultysq + \left(\relkurtpostmeanrep+2\right)\,\varsqpostmeanrep\,\cvinvsqdifficultysq/\meandifficultysq^2 +4\,\varpostmeanrep\,\cvinvsqdifficultysq\right)\,\left(1+\varpostvarrep\,\cvinvsqdifficultysq/\meandifficultysq^2\right) }}
\end{align}

For the case of evidence integration, we obtain:
\begin{align}
\perf&=\meandifficultysq/n+\varpostmeanrep\\
\corr
&=\frac{1}{\sqrt{\left(\postmeanrelkurt/n+3+\left(\postmeanrelkurt/n+2\right)\cvinvsqdifficultysq + \left(\relkurtpostmeanrep+2\right)\,\varsqpostmeanrep\,n^2\,\cvinvsqdifficultysq/\meandifficultysq^2 +4\,\varpostmeanrep\,\cvinvsqdifficultysq\right)\,\left(1+\varpostvarrep\,n^2\,\cvinvsqdifficultysq/\meandifficultysq^2\right) }}
\end{align}


\newpage
\appendix

\section{Convergence of the posterior kurtosis}

\newcommand{\ordo}[1]{\fun{\mathcal{O}}{#1}}

In this section our aim is to derive, in the context of the evidence integration model, how the kurtosis of the posterior, $\postnrelkurt$, converges as the number of data points, $n$, increases, in particular what is the leading order (in $n^{-1}$) of this convergence (at least asymptotically, in the $n\rightarrow\infty$ limit).

Based on \citep{johnson70}, the $r$th moment of the rescaled posterior distribution  (of $\bar{x}=b\,\left(x-a\right)$\footnote{The exact values of $a$ and $b$ will be unimportant later, but more specifically, $a$ is the maximum likelihood estimate of $x$, and $b$ is related to the inverse score differential at the maximum likelihood estimate.})  can be written as
\begin{align}
\E{x}{\bar{x}^r} = \sum_{j=r}^K  \lambda_{rj} \, n^{-j/2} + \ordo{n^{-\left(K+1\right)/2}} \label{eq:momentexpansion}
\end{align}
where, in general, the constants $\lambda_{rj}$ depend on the particular data points on which the posterior is conditioned but all $\lambda_{rj}$ for which $j$ is odd are zero, and for even $r$, the leading term of the sum can be given directly as
\begin{align}
&& \lambda_{rr} &= 2^{r/2} \, \fun{\Gamma}{\left(r+1\right)/2}\,\fun{\Gamma^{-1}}{1/2} \\
\text{which for } && r&=2 && \text{ and } & r &=4 \\
\text{ yields } && \lambda_{22} &= 1 && \text{ and } & \lambda_{44} &= 3 \label{eq:momentexpansion_lambda}
\end{align}

Applying Eqs.~\ref{eq:momentexpansion} and \ref{eq:momentexpansion_lambda}
to the case of the first four moments of the rescaled posterior, and expanding the first couple of terms for each, we obtain
\begin{align}
\E{x}{\bar{x}} 
 &= \sum_{j=1}^7  \lambda_{1j} \, n^{-j/2} + \ordo{n^{-\left(7+1\right)/2}}\\
 &=  \lambda_{12} \, n^{-1} + \lambda_{14} \, n^{-2} + \lambda_{16} \, n^{-3} + \ordo{n^{-4}}\\
\E{x}{\bar{x}^2} 
&= \sum_{j=2}^7  \lambda_{2j} \, n^{-j/2} + \ordo{n^{-\left(7+1\right)/2}}\\
 &= n^{-1} + \lambda_{24} \, n^{-2} + \lambda_{26} \, n^{-3} + \ordo{n^{-4}}\\
\E{x}{\bar{x}^3} 
&= \sum_{j=3}^7  \lambda_{3j} \, n^{-j/2} + \ordo{n^{-\left(7+1\right)/2}}\\
 &= \lambda_{34} \, n^{-2} + \lambda_{36} \, n^{-3} + \ordo{n^{-4}}\\
\E{x}{\bar{x}^4}
&=\sum_{j=4}^7  \lambda_{4j} \, n^{-j/2} + \ordo{n^{-\left(7+1\right)/2}}\\
 &=  3 \, n^{-2} + \lambda_{46} \, n^{-3}+ \ordo{n^{-4}}
\end{align}
The second and fourth central moments of the rescaled posterior can then be written as
\begin{align}
\E{x}{\left(\bar{x}-\E{x}{\bar{x}}\right)^2} 
&= \E{x}{\bar{x}^2}  - \left(\E{x}{\bar{x}}\right)^2 \\
&= n^{-1} + \lambda_{24} \, n^{-2} + \lambda_{26} \, n^{-3} - \left[\lambda_{12} \, n^{-1} + \lambda_{14} \, n^{-2} + \lambda_{16} \, n^{-3}\right]^2 + \ordo{n^{-4}}\\
&= n^{-1} + \lambda_{24} \, n^{-2} + \lambda_{26} \, n^{-3} - \left[\lambda_{12}^2 \, n^{-2} + 2 \, \lambda_{12} \, \lambda_{14} \, n^{-3} \right] + \ordo{n^{-4}}\\
&= n^{-1} + \underbrace{\left( \lambda_{24} - \lambda_{12}^2\right)}_{\displaystyle \bar{\lambda}_{22}} \, n^{-2} + \underbrace{\left( \lambda_{26} - 2 \, \lambda_{12} \, \lambda_{14}\right)}_{\displaystyle \bar{\lambda}_{23}} \, n^{-3} + \ordo{n^{-4}} \\
\E{x}{\left(\bar{x}-\E{x}{\bar{x}}\right)^4} 
&= \E{x}{\bar{x}^4}  - 4\, \E{x}{\bar{x}} \, \E{x}{\bar{x}^3} + 6 \, \left(\E{x}{\bar{x}}\right)^2 \E{x}{\bar{x}^2} - 3 \, \left(\E{x}{\bar{x}}\right)^4 \\
&= 3 \, n^{-2} + \lambda_{46} \, n^{-3} - \nonumber\\
&\phantom{=} - 4\, \left[\lambda_{12} \, n^{-1} + \lambda_{14} \, n^{-2} + \lambda_{16} \, n^{-3}\right]\,\left[\lambda_{34} \, n^{-2} + \lambda_{36} \, n^{-3}\right]+\nonumber\\
&\phantom{=} + 6 \, \left[\lambda_{12} \, n^{-1} + \lambda_{14} \, n^{-2} + \lambda_{16} \, n^{-3}\right]^2 \left[n^{-1} + \lambda_{24} \, n^{-2} + \lambda_{26} \, n^{-3}\right] - \nonumber\\
&\phantom{=} - 3 \, \left[\lambda_{12} \, n^{-1} + \lambda_{14} \, n^{-2} + \lambda_{16} \, n^{-3}\right]^4 + \ordo{n^{-4}} \nonumber\\
&= 3 \, n^{-2} + \lambda_{46} \, n^{-3} - 4\, \lambda_{12} \, \lambda_{34} \, n^{-3} + 6 \, \lambda_{12}^2 \, n^{-3}  + \ordo{n^{-4}}\\
&= 3 \, n^{-2} + \underbrace{\left( \lambda_{46} - 4\, \lambda_{12} \, \lambda_{34} + 6 \, \lambda_{12}^2 \right)}_{\displaystyle \bar{\lambda}_{33}} \, n^{-3}  + \ordo{n^{-4}}
\end{align}

From these, the second and fourth central moments of the original posterior can simply be written as
\begin{align}
\E{x}{\left(x-\postmean\right)^2} &= \frac{1}{b^2} \, \E{x}{\left(\bar{x}-\E{x}{\bar{x}}\right)^2} = \frac{1}{b^2} \, \left(n^{-1} + \bar{\lambda}_{22} \, n^{-2} + \bar{\lambda}_{23} \, n^{-3} \right)  + \ordo{n^{-4}}  \label{eq:varianceexpansion}\\
\E{x}{\left(x-\postmean\right)^4} &= \frac{1}{b^4} \, \E{x}{\left(\bar{x}-\E{x}{\bar{x}}\right)^4} = \frac{1}{b^4} \, \left(3 \, n^{-2} + \bar{\lambda}_{33} \, n^{-3} \right)  + \ordo{n^{-4}}
\end{align}

The kurtosis of the posterior can then be written as
\begin{align}
\postnrelkurt&=\frac{\E{x}{\left(x-\postmean\right)^4}}{\left(\E{x}{\left(x-\postmean\right)^2}\right)^2} - 3\\
&=\frac{3 \, n^{-2} + \bar{\lambda}_{33} \, n^{-3} + \ordo{n^{-4}}}
{\left(n^{-1} + \bar{\lambda}_{22} \, n^{-2} + \bar{\lambda}_{23} \, n^{-3}\right)^2 + \ordo{n^{-4}}} - 3\\
&=\frac{3 \, n^{-2} + \bar{\lambda}_{33} \, n^{-3} + \ordo{n^{-4}}}
{n^{-2} + 2\, \bar{\lambda}_{22} \, n^{-3} + \ordo{n^{-4}}} - 3
\end{align}
Expanding the first term, which is a rational function, in $n^{-1}$, one obtains
\begin{align}
\frac{3 \, n^{-2} + \bar{\lambda}_{33} \, n^{-3} + \ordo{n^{-4}}}{n^{-2} + 2\, \bar{\lambda}_{22} \, n^{-3} + \ordo{n^{-4}}}
&= a_0 + a_1 \, n^{-1} + \ordo{n^{-2}} \\
3 \, n^{-2} + \bar{\lambda}_{33} \, n^{-3} + \ordo{n^{-4}} &= 
\left[a_0 + a_1 \, n^{-1} \right] \, \left[n^{-2} + 2\, \bar{\lambda}_{22} \, n^{-3}\right] + \ordo{n^{-4}}\\
&= a_0 \, n^{-2} + 2\, a_0 \, \bar{\lambda}_{22} \, n^{-3} + a_1 \, n^{-3} + \ordo{n^{-4}}\\
&= a_0 \, n^{-2} + \left( 2\, a_0 \, \bar{\lambda}_{22} + a_1\right) \, n^{-3} + \ordo{n^{-4}}
\end{align}
from which one can derive, by matching terms of the same order on the two sides, that 
\begin{align}
a_0 &= 3\\
a_1 &= \bar{\lambda}_{33} - 2\, a_0 \, \bar{\lambda}_{22} = \bar{\lambda}_{33} - 6 \, \bar{\lambda}_{22}
\end{align}
which means that $\postnrelkurt$ can be written as
\begin{align}
\postnrelkurt&= 3 + \left(\bar{\lambda}_{33} - 6 \, \bar{\lambda}_{22}\right) \, n^{-1}  - 3 + \ordo{n^{-2}}\\
&=\left(\bar{\lambda}_{33} - 6 \, \bar{\lambda}_{22}\right) \, n^{-1}  + \ordo{n^{-2}}
\end{align}
Thus, we have just derived that the kurtosis indeed goes to zero (because in the end, the zeroth order term in the expansion of $\postnrelkurt$ is zero), and asymptotically converges as $1/n$ (because the term in the expansion that is first order in $n^{-1}$ is non-zero). (On the way, in Eq.~\ref{eq:varianceexpansion}, we have also proven the convergence of the posterior variance, and thus $\difficultynsq$, also scales asymptotically as $1/n$.)

\end{document}